\documentclass[fleqn,usenatbib,useAMS]{mnras}

\usepackage{newtxtext,newtxmath}
\usepackage{braket}
\usepackage{natbib}

\usepackage[T1]{fontenc}

\DeclareRobustCommand{\VAN}[3]{#2}
\let\VANthebibliography\thebibliography
\def\thebibliography{\DeclareRobustCommand{\VAN}[3]{##3}\VANthebibliography}

\usepackage{graphicx}	
\usepackage{amsmath}	

\title[Unified model for CV evolution]{A unified model for the evolution of cataclysmic variables}
\author[Sarkar \& Tout]{
Arnab Sarkar$^{1}$\thanks{E-mail: as3158@cam.ac.uk}
and Christopher A. Tout$^{1}$\thanks{E-mail: cat@ast.cam.ac.uk}
\\
$^{1}$Institute of Astronomy, The Observatories, Madingley Road, Cambridge CB3 OHA, UK
}

\date{Accepted XXX. Received YYY; in original form ZZZ}

\pubyear{2015}

\begin{document}
\label{firstpage}
\pagerange{\pageref{firstpage}--\pageref{lastpage}}
\maketitle

\begin{abstract}
We give an updated version of the analytical equation of state used in the Cambridge stellar evolution code (STARS) as a free to use open-source package that we have used to model cool white dwarfs down to temperatures $\log_{10}(T_\mathrm{eff}/\mathrm{K})\:=\;3$. With this update in the STARS code we model the secular evolution of cataclysmic variable (CV) stars using a double dynamo model wherein there is an interplay between two $\alpha-\Omega$ dynamos, one in the convective envelope and the other at the boundary of a slowly rotating shrinking radiative core and the growing convective envelope. We confirm that this model provides a physical formalism for the interrupted magnetic braking paradigm. In addition, our model also provides a mechanism for extra angular momentum loss below the period gap. We construct the relative probability distribution of orbital periods $P_\mathrm{orb}$ using the {mass} distribution of white dwarfs in cataclysmic variables and find that our model excellently reproduces the period gap and the observed period minimum spike in CV distribution. We also compare the evolutionary trajectories from our model with those of other empirical models and find agreement between the two. We also report good agreement between our modelled systems and observational data.
\end{abstract}


\section{Introduction} 
Cataclysmic variables are a class of interacting binary systems consisting of a mass-transferring secondary star along with a mass-accreting white dwarf (WD) primary \citep[][]{2003cvs..book.....W}. The secular evolution of CVs is driven by the loss of angular momentum, which leads to the secondary filling its Roche lobe and commencing mass transfer. According to the canonical model of CV evolution, for longer orbital periods ($P_\mathrm{orb}\gtrsim 3\,\mathrm{hr}$) the primary mode of angular momentum loss is some sort of magnetic braking (MB) owing to a stellar wind from the donor star. A dearth of observed mass transferring CVs between $2\,\lesssim  P_\mathrm{orb}/\mathrm{hr}\lesssim3$ (called the period gap) led to the interrupted magnetic braking paradigm \citep[][]{1983ApJ...275..713R} wherein MB stops abruptly when the donor becomes fully convective (at $P_\mathrm{orb}\approx 3\,\mathrm{hr}$). While transferring mass the donor had been driven out of thermal equilibrium. At this point it begins to regain thermal equilibrium and contracts within its Roche lobe causing the cessation of mass transfer. From here on only gravitational radiation remains as a mechanism for angular momentum loss. Mass transfer begins again only when the Roche lobe catches up with the convective donor at $P_\mathrm{orb}\approx2\,\mathrm{hr}$. The evolution of CVs is also governed by the interplay between the donor's mass-loss timescale $\tau_\mathrm{ML} \approx M_2/\Dot{M}_2$ and its Kelvin-Helmholtz or thermal timescale $\tau_\mathrm{KH}\approx GM_2^2/R_\ast L_\ast$, where $M_2$, $R_\ast$ and $L_\ast$ are the donor's mass, radius and luminosity. As long as $\tau_\mathrm{ML} \gg \tau_\mathrm{KH}$, the donor is able to maintain thermal equilibrium and behave like a standard main-sequence star. However, when $\tau_\mathrm{ML} \approx \tau_\mathrm{KH}$ mass transfer leads to an increase in the donor's size and $P_\mathrm{orb}$ increases in response to it. This leads to a period minimum $P_\mathrm{min}$ as the donor transforms from a shrinking MS donor to an expanding, partially degenerate donor \citep[][]{1981ApJ...248L..27P, 1982ApJ...254..616R}. 

Theoretical predictions of $P_\mathrm{min}$ have often disagreed with observations. Initial calculations showed $P_\mathrm{min}\approx 65 \,\mathrm{min}$ which was substantially shorter than the then observed cutoff of $P_\mathrm{min}\approx 75 \,\mathrm{min}$ \citep{Knigge2006}. Later, an SDSS sample of intrinsically faint CVs found  $80\lesssim P_\mathrm{min}/\mathrm{min}\lesssim86$ making the discrepancy even greater. This means that there should be a mechanism for the loss of angular momentum below the period gap in addition to gravitational radiation. Although \cite{Knigge2011} show that multiplying the gravitational radiation angular momentum loss term by $2.47\pm0.22$ is able to reproduce the extra angular momentum loss (AML) below the period gap, a physical mechanism is yet to be established. Similarly, many groups have modelled CVs above the period gap using an empirical formula for MB given by \cite{1983ApJ...275..713R} which, along with gravitational radiation, is able to reproduce the desired period gap in the trajectory of a particular system. However, a physically motivated MB mechanism is yet to be established as well as a sensitivity analysis of CVs to the WD mass, novae etc. Some have come up with dynamo models wherein magnetic fields are suppressed when the donor becomes fully convective  \citep[see][]{Charbonneau1997, Zangrilli1997}. These can potentially model this braking mechanism. It is noted that the complete cessation of angular momentum loss owing to the suppression of stellar magnetic fields at the period gap is not entirely correct because there is evidence of remnant stellar magnetism in fully convective stars \citep[see the discussion by][]{Knigge2011}. We argue that this remnant can account for a possible AML mechanism below the period gap. However, in order to model the evolution of CVs near and below the period gap, it is essential to have a robust equation of state implemented in stellar evolution codes because the surface temperature of the donor falls to $\mathrm{log_{10}}(T_\mathrm{eff}/\mathrm{K})\lesssim3.5$. At these temperatures and for matter densities $\mathrm{log_{10}}(\rho/\mathrm{g\,cm^{-3}})\lesssim1$, pressure ionization makes a significant contribution to the total pressure but the exact behaviour of matter in this region is still not well understood\footnote{See for instance the stellar evolution code MESA's equation of state implementation in the $(\rho,\,T)$ plane (in \url{https://docs.mesastar.org/en/latest/eos/overview.html}) which uses interpolation between various component equation of state modules.}. In this paper we improve on the equation of state module in the Cambridge stellar evolution code, STARS, and use this with a revised double dynamo (DD) model of \cite{Zangrilli1997} to explain the interrupted magnetic braking mechanism.

In section \ref{eos} we use the updated equation of state module from STARS to create models of cool white dwarfs with surface temperatures $T_\mathrm{eff}=10^3\,\mathrm{K}$ and compare our results to those of the HELM equation of state. We explain the DD model with all its revision in detail in section \ref{DD}. Using the updated STARS code, we model the secular evolution of zero-age cataclysmic variables with the revised DD model in section \ref{secev}. In section \ref{results} we present the results from our modelled systems and compare them with the work done by others and with observational data. We summarize our results in section \ref{Conclusion}.
\section{The STARS equation of state package}
\label{eos}

The mathematical formalism of the STARS equation of state (EOS) has been given by \cite{1973A&A....23..325E} and \cite{Pols1995}. We modify the EOS {in order to model the evolution of the donor close to the period minimum, where the secondary is a semi-degenerate, cool star with $\mathrm{log_{10}}(T_\mathrm{eff}/\mathrm{K})\approx3$. We do this} by correcting the implementation of the number of $\mathrm{H_2}$ molecules per unit mass $N_\mathrm{H_2}$. Otherwise this became undefined in regions where $X_\mathrm{H} =0$ causing our EOS module to crash. {For instance, $N_\mathrm{H_2}$ in equation 13 of \cite{Pols1995} and hence all its associated derivatives ($\partial N_\mathrm{H_2}/\partial \mathrm{ln}f$ and $\partial N_\mathrm{H_2}/\partial \mathrm{ln}T$) became undefined as $N_\mathrm{H}\rightarrow0$ \citep[see][for the definition of $f$]{1973A&A....23..325E}. This is resolved in our current update.} We also make corrections to the derivatives of the compensation term $\Delta F_\mathrm{PI}(N_\mathrm{e0},V,T)$ \citep[see section 2.2.2, particularly equation 28 of][]{Pols1995} {where we define the derivatives $\partial \rho_{\ast0}/\partial \mathrm{ln}f$ and $\partial \rho_{\ast0}/\partial \mathrm{ln}T$ as\begin{center}
\begin{equation}
\label{del1}
\frac{\partial \rho_{\ast0}}{\partial \mathrm{ln}f} = \frac{\partial \rho_\ast}{\partial \mathrm{ln}f} - \frac{1}{N_\mathrm{e0}}\frac{\partial N_\mathrm{e}}{\partial \mathrm{ln}f}
 \end{equation}
\end{center}
and\begin{center}
\begin{equation}
\label{del2}
\frac{\partial \rho_{\ast0}}{\partial \mathrm{ln}T} = \frac{\partial \rho_\ast}{\partial \mathrm{ln}T} - \frac{1}{N_\mathrm{e0}}\frac{\partial N_\mathrm{e}}{\partial \mathrm{ln}T},
 \end{equation}
\end{center}
where $\rho_\ast$ and $N_\mathrm{e}$ are defined by \cite{1973A&A....23..325E} and \cite{Pols1995}, $\rho_{\ast0}$ is the corresponding Fermi-Dirac integral for the compensation term $\Delta F_\mathrm{PI}(N_\mathrm{e0},V,T)$ and $N_\mathrm{e0}$ is the total number of electrons assuming complete ionization of all species. With these changes our code works well down to temperatures $T\approx10^3\,\mathrm{K}$.} 

\subsection{Comparison with HELM EOS}
\begin{figure}
\includegraphics[width=0.5\textwidth]{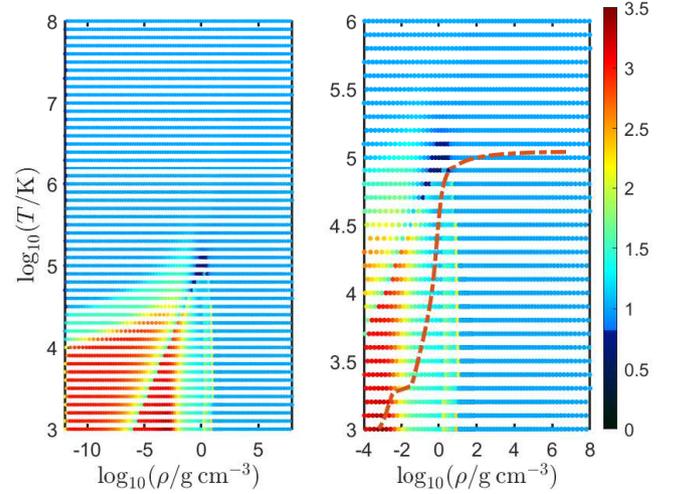}
\caption{The ratio $P_\mathrm{g,{STARS}}/P_\mathrm{g,{HELM}}$ plotted in the $(\rho,T)$ plane for $X_\mathrm{H}=0.0$ and $Z=0.02$ shown for two different but overlapping temperature and density ranges. The orange line is the run of temperature with density of a $1M_\odot$ $\mathrm{He}$ WD with $T_\mathrm{eff} = 10^3\:\mathrm{K}$. }
\label{fig:eos}
\end{figure}
We compare our EOS with the HELM equation of state \citep[][]{2000ApJS..126..501T} in the $(\rho,T)$ plane in the region $-12\leq\log_{10}(\rho/\mathrm{g\,cm^{-3}})\leq8$ and $3\leq\log_{10}(T/\mathrm{K})\leq8$ in Fig.~\ref{fig:eos}. We find that the gas pressures ($P_\mathrm{g}$) of both the EsOS agree very well with each other over the entire $(\rho,T)$ plane, except the region $-12\leq\log_{10}(\rho/\mathrm{g\,cm^{-3}})\leq1$ and $3\leq\log_{10}(T/\mathrm{K})\leq4.5$ where pressure ionization contributes significantly to the gas pressure $P_\mathrm{g}$. The ratio is $0.5\lesssim P_{\mathrm{g},\mathrm{STARS}}/P_{\mathrm{g},\mathrm{HELM}}\lesssim3$ overall. There is still some multi-valuedness in the region $\log_{10}(\rho/\mathrm{g\,cm^{-3}})\approx1$ and $\log_{10}(T/\mathrm{K})\leq4.5$ \citep[see section 2.3 and Fig.~1 of][]{Pols1995}, although our white dwarf (WD) tracks in the $(\rho,T)$ plane never cross this discontinuity. To represent this we also plot the EOS track of a $1M_\odot$ $\mathrm{He}$ WD with $T_\mathrm{eff}=10^3 \mathrm{\:K}$. We see that only the photosphere of the WD lies on the region where there is some considerable difference in $P_\mathrm{g}$ between STARS and HELM. However, it is important to mention that even the HELM EOS does not have robust estimates for gas pressures in this region. Our EOS module can be found as an open-source package at \url{https://github.com/ArnabSarkar3158/STARS-EOS-}.

\section{The double dynamo model}
\label{DD}
The secular evolution of CVs driven by the combination dynamo along with expressions for angular momentum loss has been explained for a bipolytropic model by \cite{Zangrilli1997}, hereinafter ZTB, where the dynamo itself has been explained by \cite{Tout1992}, hereinafter TP, for a fully convective star. Here we revisit the equations and concepts that we incorporate in the STARS code. {We also revise the current DD model and construct a tuple of three physically motivated free parameters $(\alpha,\beta,\gamma)$ with the aim of {constructing a formalism for the secular evolution of the donor star in CVs} that can explain the period gap, the observed period minimum spike and an additional angular momentum loss mechanism below the period gap in CVs.}

\subsection{The envelope dynamo}
\label{convDD}
We assume that the convective envelope of the donor is tidally locked with the orbit of the system and corotates with it so its angular velocity is\begin{center}
\begin{equation}
\label{orb_eq}
\Omega_\mathrm{env} = \Omega_\mathrm{orb} = \Omega.
 \end{equation}
\end{center}
We further assume that magnetic fields are created and destroyed at the same rate and use the equilibrium equations (2) and (3) of ZTB,
\begin{center}
\begin{equation}
\label{dyn1}
\frac{\mathrm{d}B_\mathrm{\phi}}{\mathrm{d}t} =\Delta\Omega B_\mathrm{p} - \frac{B_\mathrm{\phi}}{\tau_\mathrm{\phi}} = 0
\end{equation}
\end{center}
and
\begin{center}
\begin{equation}
\label{dyn2}
\frac{\mathrm{d}A_\mathrm{\phi}}{\mathrm{d}t} =\Gamma B_\mathrm{\phi} - \frac{A_\mathrm{\phi}}{\tau_\mathrm{p}} = 0,
 \end{equation}
\end{center}
where $B_\mathrm{\phi}$ is the toroidal component of the magnetic field in the donor, $A_\mathrm{\phi}$ is the azimuthal component of the magnetic vector potential, such that the poloidal component of the magnetic field $B_\mathrm{p}\approx A_\mathrm{\phi}/R_\ast$ can be defined and $\tau_\mathrm{\phi}$ and $\tau_\mathrm{p}$ are the time-scales on which the polidal and toroidal magnetic field components are destroyed. Equation (\ref{dyn2}) can then be written as
\begin{center}
\begin{equation}
\label{dyn3}
\frac{\mathrm{d}B_\mathrm{p}}{\mathrm{d}t} =\frac{\Gamma}{R_\ast} B_\mathrm{\mathrm{\phi}} - \frac{B_\mathrm{p}}{\tau_\mathrm{p}} = 0,
 \end{equation}
\end{center}
where $\Gamma$ is the regeneration term (also known as the $\alpha$ term in $\alpha-\Omega$ dynamo model) and $\Delta\Omega$ is the shear term (or the $\Omega$ term in the $\alpha-\Omega$ dynamo model) which corresponds to a measure of differential rotation, to be discussed in the next section. We also note that we expect the shear term to act more rapidly than the regeneration term and therefore expect $B_\mathrm{\phi}\gg B_\mathrm{p}$ such that $B_\mathrm{p}\approx \epsilon B_\mathrm{\phi}$ where $\epsilon\ll 1$. We now use ZTB's expression for energy input into the wind\begin{center}
\begin{equation}
\label{Lw_conv}
L_{\mathrm{w,conv}} \approx \frac{1}{1800}\frac{M_\mathrm{env}}{M_2}\left(\frac{R_\ast}{R_\mathrm{env}}\right)^2L_\ast,
\end{equation}
\end{center}
where $R_\ast,\:L_\ast,\:M_\mathrm{env}$ and $R_\mathrm{env}$ are the donor's radius, luminosity, mass in the convective envelope and radius in the convective envelope. We then use equation (2.4) of TP for the convective envelope, \begin{center}
\begin{equation}
\label{Lw_conv2}
L_{\mathrm{w,conv}} \approx \frac{GM_\mathrm{env}\Dot{M}_\mathrm{w,conv}}{R_\ast}.
\end{equation}
\end{center}
And with equations (\ref{Lw_conv}) and (\ref{Lw_conv2}) we arrive at an expression for the mass-loss rate owing to the convective dynamo
\begin{center}
\begin{equation}
\label{mlconv}
\Dot{M}_\mathrm{w,conv} \approx \left(\frac{M_\mathrm{conv}}{M_2}\right)^\alpha\frac{1}{1800} \frac{R_\ast}{GM_\mathrm{env}} \frac{M_\mathrm{env}}{M_2}\left(\frac{R_\ast}{R_\mathrm{env}}\right)^2 L_\ast,
\end{equation}
\end{center}
where $M_\mathrm{conv}$ is the mass at which the star becomes fully convective. The term $(M_\mathrm{conv}/M_2)^\alpha$ acts as an enhancement to the efficiency of the convective mass loss when the donor is fully convective and $\alpha$ is one of the three free parameters in our DD model. We attribute this to two processes. First is differential rotation in the fully convective star. Models of single stars with mass around $M_\odot$ suggest that there is strong differential rotation between the core and the surface with a flat rotation rate in the external part and an increased rotation in the stellar core \citep[][]{1996ApJ...466.1078R, Eggenberger2019}. However, our model for the donor star assumes that the rotation of the convective envelope is fast compared to that of the core (see the next subsection). This is a valid assumption until the donor becomes fully convective and the whole star corotates with the orbit. So we need to take into account differential rotation for a fully convective star. A higher rotation rate at the centre leads to a higher energy input into the wind through higher order correction terms (equation 6 of ZTB) and consequently enhances the mass-loss rate of the envelope dynamo. Secondly we consider an additional viscosity to account for extra angular momentum transport in low-mass stars \citep[][]{Dumont_slides}, and required for stability in a system under gravitational-convective perturbations \citep[][]{Snytnikov2011}. Increased viscosity again leads to an increased wind energy input which enhances the convective mass loss. As explained by ZTB, because the convective dynamo only influences the evolution below the period gap, we set $M_\mathrm{conv}\approx0.2 M_\odot$ until our system reaches the upper end of the period gap, after which the code computes $M_\mathrm{conv}$.

\subsection{The boundary layer dynamo}
ZTB's model assumes that, unlike the convective envelope, the radiative core is neither tidally spun up nor linked to the envelope magnetically. Therefore we can assume that $\Omega_\mathrm{core}\approx0$. Thus at the boundary layer
\begin{center}
\begin{equation}
\label{orb_eq2}
\Delta\Omega = \Omega_\mathrm{env}-\Omega_\mathrm{core} \approx \Omega.
\end{equation}
\end{center}
We use the expression for wind luminosity of ZTB (their equation~22 multiplied by 0.1 to account for the fact that the energy input is inefficient)
\begin{center}
\begin{equation}
\label{Lw_bl}
L_{\mathrm{w,bl}} \approx 10\gamma_\mathrm{c} v_\mathrm{c} \left(\frac{R_\mathrm{env}}{R_\ast}\right)\Delta\Omega^2H_\mathrm{B}^2R_\mathrm{core}^2\cdot4\pi\rho_\mathrm{B},
\end{equation}
\end{center}
where $\gamma_\mathrm{c}\approx10^{-2}$ is the efficiency of the $\Gamma$ regeneration term (see TP for a thorough discussion), $R_\mathrm{core}$ is the radius of the radiative core, $\rho_\mathrm{B}$ is the density at the boundary layer, $v_\mathrm{c}\approx (L_\ast R_\ast/\eta M_2)^{1/3}$ is the convective velocity, the constant $\eta\approx30$ \citep[see][]{Campbell1983} and 
\begin{center}
\begin{equation}
\label{Hb}
H_\mathrm{B} = 0.001\mathrm{min}\left(H_\mathrm{p},\frac{1}{2}R_\mathrm{env}\right)
\end{equation}
\end{center}
is the thickness of the boundary layer, with $H_\mathrm{p}$ the pressure scale-height at the boundary layer. We approximate the local gravity at the boundary layer by
\begin{center}
\begin{equation}
\label{gb}
g_\mathrm{B} = \frac{4}{3}\pi G \braket{\rho} R_\mathrm{core},
\end{equation}
\end{center}
where $\braket{\rho}$ is the average density of the donor\footnote{We have also tried using $g_\mathrm{B} = G(M_2-M_\mathrm{env})/R_\mathrm{core}^2$ which gives the same final results as equation~(\ref{gb}). However this expression is not well behaved when the core vanishes.}. We again use equation~(2.4) of TP for the boundary layer
\begin{center}
\begin{equation}
\label{Lw_bl2}
L_{\mathrm{w,bl}} \approx \frac{GM_\mathrm{B}\Dot{M}_\mathrm{w,bl}}{R_\mathrm{core}}
\end{equation}
\end{center}
where $M_\mathrm{B} = \rho_\mathrm{B}\cdot4\pi R_\mathrm{core}^2H_\mathrm{B}$ is the mass of the boundary layer. Using equations (\ref{Lw_bl}) and (\ref{Lw_bl2}) we arrive at
\begin{center}
\begin{equation}
\label{mlbl}
\Dot{M}_\mathrm{w,bl} \approx \beta \frac{R_\mathrm{core}}{GM_\mathrm{B}}10\gamma_\mathrm{c} v_\mathrm{c} \left(\frac{R_\mathrm{env}}{R_\ast}\right)\Delta\Omega^2H_\mathrm{B}^2R_\mathrm{core}^2\cdot4\pi\rho_\mathrm{B}.
\end{equation}
\end{center}
Here $\beta$ is the efficiency with which mass loss is driven by the boundary layer dynamo. We make $\beta$ one of the three free parameters in our DD model.

\subsection{The combination dynamo}

 Importantly we assume that the magnetic field in the wind is predominantly the poloidal component of the donor's stellar field and model $B_\mathrm{p,conv}$ with equation (4.10) of TP,
\begin{center}
\begin{equation}
\label{bpconv}
B_\mathrm{p,conv} = 10\gamma_\mathrm{c} v_\mathrm{c}\sqrt{4\pi \rho_\mathrm{conv}},
\end{equation}
\end{center}
where $\rho_\mathrm{conv}$ is the density of the convective envelope such that
\begin{center}
\begin{equation}
\label{rhoconv}
\rho_\mathrm{conv} = \frac{M_\mathrm{env}}{4\pi R_\mathrm{core}^2R_\mathrm{env}}.
\end{equation}
\end{center}
However we ensure that $\rho_\mathrm{conv}$ remains well defined when the radiative core vanishes by letting $\rho_\mathrm{conv}$ tend to $\braket{\rho}$. Similarly, we model $B_\mathrm{p,bl}$ with
\begin{center}
\begin{equation}
\label{bpbl}
B_\mathrm{p,bl} = v_\mathrm{p}\sqrt{4\pi \rho_\mathrm{B}},
\end{equation}
\end{center} where $\rho_\mathrm{B}$ is the density at the boundary and $v_\mathrm{p}$ is the poloidal Alfvén speed given by equation (20) of ZTB,
\begin{center}
\begin{equation}
\label{vp}
v_\mathrm{p} \approx 10\gamma_\mathrm{c}^{2/3}v_\mathrm{c}^{2/3}\left(\frac{R_\mathrm{env}}{R_\ast}\right)^{2/3}\Delta\Omega^{1/3}H_\mathrm{B}^{1/3}.
\end{equation}
\end{center} The total wind mass-loss rate of the donor is
\begin{center}
\begin{equation}
\label{mltot}
\Dot{M}_\mathrm{w} = \Dot{M}_\mathrm{w,conv} + \Dot{M}_\mathrm{w,bl}
\end{equation}
\end{center}and its total poloidal magnetic field is\begin{center}
\begin{equation}
\label{bptot}
B_\mathrm{p} = B_\mathrm{p,conv} + B_\mathrm{p,bl}.
\end{equation}
\end{center}

\subsection{Angular momentum loss prescription}
The rate of loss of angular momentum from the donor in the wind is given by
\begin{center}
\begin{equation}
\label{jdot}
\Dot{J}_\mathrm{w} = -\Dot{M}_\mathrm{w}\Omega R_\mathrm{A}^2,
\end{equation}
\end{center}
where $R_\mathrm{A}$ is the Alfvén radius of the donor. We model this assuming that the magnetic field is dipolar such that (see section 2 of TP for a thorough derivation),
\begin{center}
\begin{equation}
\label{ra}
R_\mathrm{A} = R_\ast \left(\frac{B_\mathrm{p}^2R_\ast^2}{\Dot{M}_\mathrm{w}v_\mathrm{w}}\right)^{1/4},
\end{equation}
\end{center}
where $v_\mathrm{w} = \sqrt{2GM_2/R_\ast}$ is the escape velocity of the donor. Because the donor is tidally locked with the orbit, loss of angular momentum from the donor leads to loss of orbital angular momentum. Finally, writing the orbital angular momentum of the system as 
\begin{center}
\begin{equation}
\label{j}
{J}= \frac{M_1M_2}{M_1+M_2}\Omega a^2,
\end{equation}
\end{center}
where $M_1$ is the mass of the WD accretor and $a$ is the orbital separation, we arrive at 
\begin{center}
\begin{equation}
\label{jdotbyj}
\frac{\Dot{J}_\mathrm{w}}{J} = -\left(1+ \frac{R_\mathrm{env}}{R_\ast}\right)^\gamma \Dot{M}_\mathrm{w}\frac{M_1+M_2}{M_1M_2}\left(\frac{R_\mathrm{A}}{a}\right)^2,
\end{equation}
\end{center}
where the term $(1 + R_\mathrm{env}/R_\ast)^\gamma$ is the efficiency with which angular momentum is lost from the donor. This efficiency is of the order of unity for a thin convective envelope and $2^\gamma$ for a fully convective donor. It acts to increase the Alfvén radius of the donor when $M_2\ll1M_\odot$. An increased $R_\mathrm{A}$ has been postulated in M dwarf stars by \citet[~table 1]{VillarrealDAngelo2017}. It can be attributed to the fact that the Alfvén radius is sensitive the density of outflows, and low-density outflows lead to $R_\mathrm{A}\gg R_\ast$ (see section 2 of TP). Our donor passes through a phase where it closely resembles an M dwarf star so we argue that it must experience a rapid increase in its Alfvén radius during its transition from a solar-like star to M dwarf-like. We use $\gamma$ as the third free parameter in our DD model.

\section{Secular evolution of cataclysmic variables using the double dynamo model}
\label{secev}

With our updated STARS code, we model the secular evolution of zero-age cataclysmic variables (ZACVs) wherein Roche lobe overflow (RLOF) begins from a zero-age main-sequence (ZAMS) donor. Throughout the rest of this work we evolve a donor star $M_2 = 1M_\odot$\footnote{{We show in appendix~\ref{app:1} that ZACVs with donors of different masses follow the same evolutionary trajectory as long as the donor has undergone no nuclear evolution.}}, starting from a detached phase when there is no mass transfer until the donor star becomes partly degenerate at its period minimum in the $(M_2,P_\mathrm{orb})$ plane \citep[][]{1982ApJ...254..616R, 1981ApJ...248L..27P}. We assume that the mass transfer is fully non-conservative, meaning that all the mass accreted on to the WD, $M_1$ or $M_\mathrm{WD}$ used interchangeably, is expelled in the form of nova eruptions and carries away specific angular momentum of the {WD accretor}. We start with a detached system with $P_\mathrm{orb}\:=\:12\,\mathrm{hr}$ with initial period below the bifurcation limit \citep[see][for a thorough discussion]{1988A&A...191...57P, Podsiadlowski2003}, which we take to be around $P_\mathrm{orb}\:\approx\:22\,\mathrm{hr}$ for a $1M_\odot$ star\footnote{{We have not attempted to find the true bifurcation period for our donor in this work. It is very sensitive to the assumed AML mechanism.}} \citep[see equation 1 of][]{Kalomeni2016}, with no nuclear evolution of the donor. Along with the angular momentum loss prescription, equation (\ref{jdotbyj}), we add the angular momentum loss expression owing to gravitational radiation \citep[see][and the references therein]{1981ApJ...248L..27P}
\begin{center}
\begin{equation}
\label{jdotbyjgr}
\frac{\Dot{J}_\mathrm{gr}}{J} = -32G^3{{M_1M_2}}\frac{M_1+M_2}{5a^4c^5}.
\end{equation}
\end{center}
We show in sections \ref{pg} and \ref{aml} how our model reproduces the period gap and accounts for the extra AML below the period gap.

\subsection{The period gap}
\label{pg}
{The period gap is considered to be the region $2\lesssim P_\mathrm{orb}/\mathrm{hr}\lesssim3$ where there is an observed dearth of non-magnetic (or weakly magnetic) semi-detached or mass-transferring CVs, although there are systems that have been found with $P_\mathrm{orb}$ in this region because it is possible for a CV to be born with an initial period between $2$ and $3\,\mathrm{hr}$. {\protect\cite{Knigge2006} has determined the period gap to be between $P_\mathrm{orb,pg,lower}\:=\:2.15\pm0.03\,\mathrm{hr}$ and $P_\mathrm{orb, pg,upper}\:=\:3.18\pm0.04\,\mathrm{hr}$, and we use this to calibrate our models in this work.} Other systems that have been found in the period gap include polars, in which the WDs have very strong magnetic fields, and AM~CVn's which are binaries where the donor is hydrogen-exhausted \citep[see][for a review of these candidates]{2003cvs..book.....W}.
In order to reproduce the period gap our AML mechanism owing to the boundary layer dynamo needs to turn off at $P_\mathrm{orb}\approx3\,\mathrm{hr}$, causing the cessation of mass transfer. Mass transfer should begin at $P_\mathrm{orb}\approx2\,\mathrm{hr}$ when the donor regains thermal equilibrium. We show that this can be well modelled by our DD model by adjusting the free parameters $\beta$ and $\gamma$.} 
\subsubsection{The dependence on $\beta$ and $\gamma$}

\begin{figure}
\includegraphics[width=0.5\textwidth]{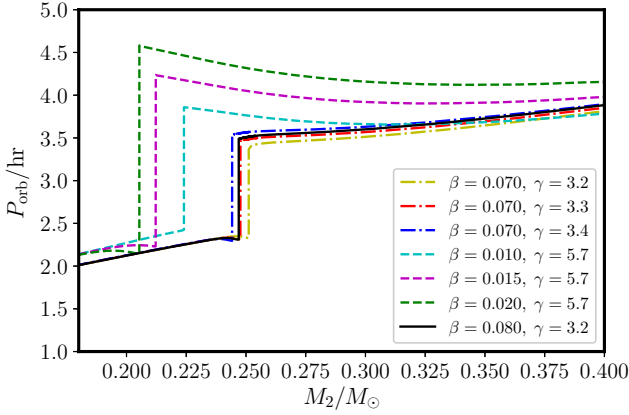}
\caption{Evolutionary tracks of our CV model with different $\beta$ and $\gamma$ choices in the $(M_2,P_\mathrm{orb})$ plane for $M_2 = 1M_\odot$ and $M_1=0.83M_\odot$. The evolution is in the direction of decreasing mass and the vertical section of the trajectory when $M_2$ is constant indicates the period gap. The dashed and dash-dotted lines are various combinations of $\beta$ and $\gamma$ which demonstrate the behaviour of the system to changes in $\beta$ and $\gamma$ respectively. As can be seen, both higher $\beta$ and $\gamma$ lead to a wider period gap and lower $M_\mathrm{conv}$. Therefore $\beta$ and $\gamma$ need to be calibrated to reproduce the period gap. The solid line is the tuple $(\beta,\gamma) =(0.08,3.2)$ that we use in further calculations.}
\label{fig:pg_pm}
\end{figure}

We first set $\alpha=0$ in equation (\ref{mlconv}) and evolve our $1M_\odot$ donor with a WD accretor of mass $M_1=0.83M_\odot$. We show later that the period gap varies with $M_1$. However we make this particular choice for $M_1$ as suggested by \cite{Pala2020} for the average WD mass in CVs \citep[see also fig.~6 of][]{Wijnen2015}. We show below that the choices of the free parameters $\beta$ and $\gamma$ influence the period gap. Fig.~\ref{fig:pg_pm} shows the period gap in the $(M_2,P_\mathrm{orb})$ plane for various combinations of $\beta$ and $\gamma$. We see that changing $\beta$ or $\gamma$ affects both the size of the period gap and the donor's mass when it becomes fully convective ($M_\mathrm{conv}$). In general $\Dot{J}_\mathrm{W}/J \propto \sqrt{\beta}$ when the boundary layer dynamo is dominant. Thus changing $\beta$ changes the angular momentum loss rate by a constant. On the other hand changing $\gamma$ leads to a more dynamic change in $\Dot{J}_\mathrm{W}/J$ because the scaling varies from about $1$ to $2^\gamma$ as a function of the fraction of the convective region of the star. We find that using $\beta=0.08$ and $\gamma=3.2$ (the solid black curve in Fig.~\ref{fig:pg_pm}) gives us the best coverage of the period gap and we fix these in section \ref{wdmass}.

\subsubsection{The dependence on WD mass}
\label{wdmass}
\begin{figure}
\includegraphics[width=0.5\textwidth]{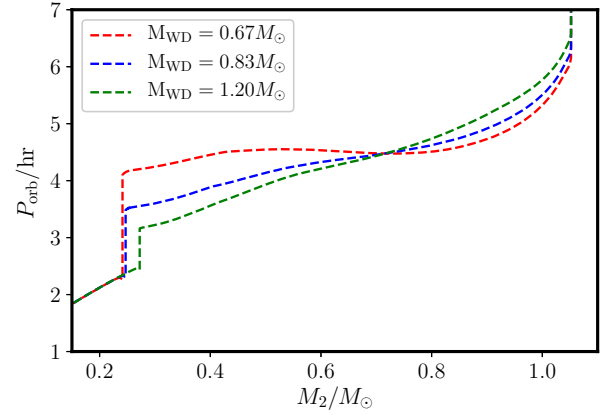}
\caption{Evolutionary tracks of CV models for different WD masses $M_1$, when $\beta=0.08$ and $\gamma= 3.2$.}
\label{fig:pg_pm_wd}
\end{figure}

Fig.~\ref{fig:pg_pm_wd} shows the dependence of the period gap and the trajectory on the WD mass in the $(M_2,P_\mathrm{orb})$ plane. Our choice of WD mass is motivated by the distribution of WDs in CVs \citep[see fig.~6 of][]{Wijnen2015}. We find that less massive WDs lead to a longer upper boundary of the period gap and a smaller mass of the donor when it becomes fully convective. This can be explained by the fact that less massive accretors drive higher mass-transfer rates and so the donor is driven more out of thermal equilibrium leading to a longer upper boundary of the period gap. Similarly, a higher mass-transfer rate means that the donors have less time to become fully convective, leading to lower $M_\mathrm{conv}$. It can be seen that the period gap is sensitive to the choice of $M_1$.

\subsection{Extra angular momentum loss (AML) below the period gap}
\label{aml}
\begin{figure}
\includegraphics[width=0.5\textwidth]{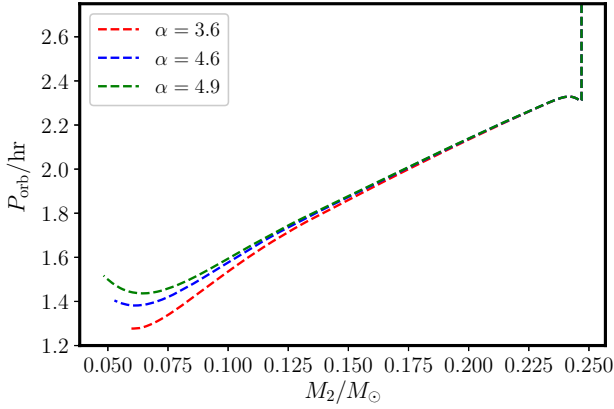}
\caption{Evolutionary tracks of CV model for different $\alpha$, when $\beta=0.08$ and $\gamma= 3.2$.}
\label{fig:pg_aml}
\end{figure}

We explained in section \ref{convDD} that the term containing $\alpha$ does not influence the evolutionary track of the CV above the period gap because the boundary layer dynamo dominates there. We now evolve our CV from the lower end of the period gap with varying $\alpha$, keeping $\beta = 0.08$ and $\gamma=3.2$. Fig.~\ref{fig:pg_aml} shows the evolutionary track for three different $\alpha$s, which were chosen to reproduce various $P_\mathrm{orb, min}$. We find $\alpha=3.6$ corresponds to $P_\mathrm{orb, min}\:=\:76.6\,\mathrm{min}$, the minimum period predicted by \cite{Knigge2006}, while $\alpha=4.6$ and $\alpha=4.9$ correspond to the more recent estimates of $P_\mathrm{orb, min}\:=\:82.89\,\mathrm{min}$ and $P_\mathrm{orb, min}\:=\:86.19\,\mathrm{min}$. These lie in the period minimum spike of \cite{Gnsicke2009}. We select $\alpha = 4.6$ for further calculations.

\section{Results}
\label{results}
{With a physical calibration of our choice of the free parameters $(\alpha, \beta,\gamma)$ in the DD model, we are now in a position to evolve ZACVs from the beginning of mass-transfer to beyond the period minimum. In section \ref{ss:evtrack} we present complete evolutionary tracks of ZACVs and explain in detail the evolution of important terms in the DD model. In section \ref{ss:pr} we construct relative probability distributions of observing a semi-detached CV with a given orbital period and thence attempt to reproduce the period gap and the period minimum spike in CV distribution. In section \ref{ss:otherauth} we compare our model with the work by \cite{Knigge2011} and in section \ref{ss:obs} we compare our results with observed CVs in the $(M_2,P_\mathrm{orb})$ plane.}

\subsection{Complete evolutionary tracks}
\label{ss:evtrack}
\begin{figure}
\includegraphics[width=0.5\textwidth]{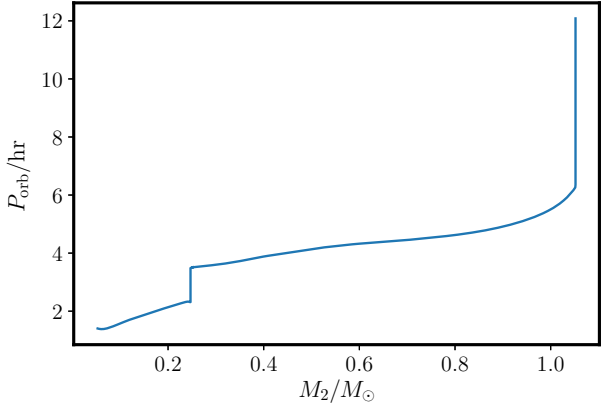}
\caption{Full evolutionary track of our CV model in the $(M_2,P_\mathrm{orb})$ plane, where $(\alpha,\beta,\gamma) = (4.6,0.08,3.2)$ and $M_1=0.83M_\odot$. This CV enters the period gap between 2.3 and 3.4 $\mathrm{hr}$ and it reaches a period minimum of $\:82.89\,\mathrm{min}$. }
\label{fig:pmfull}
\end{figure}

\begin{figure}
\includegraphics[width=0.5\textwidth]{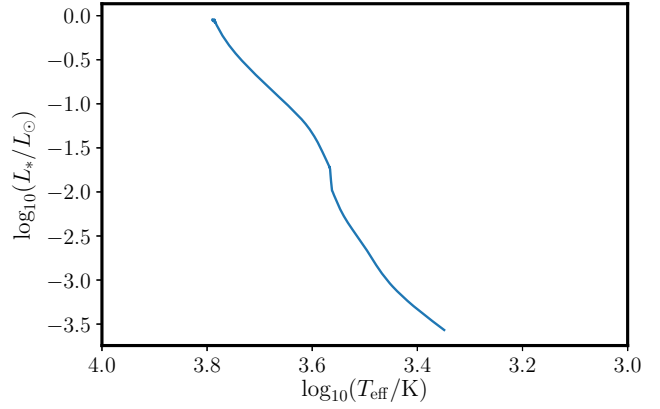}
\caption{The evolution in an HR diagram of the donor in Fig.~\ref{fig:pmfull}.}
\label{fig:hr}
\end{figure}

\begin{figure}
\includegraphics[width=0.5\textwidth]{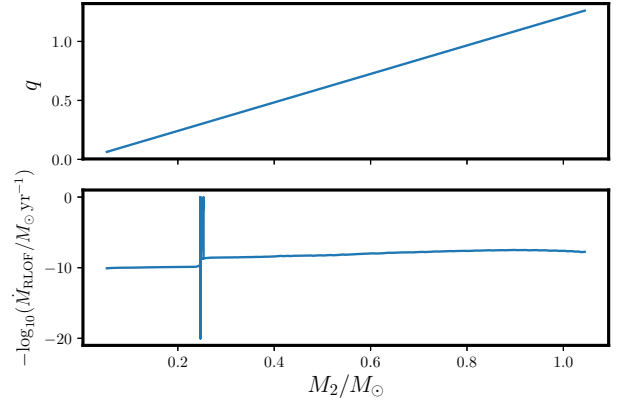}
\caption{The mass ratio $q$ and mass-transfer rate $\Dot{M}_\mathrm{RLOF}$ of the system in Fig.~\ref{fig:pmfull} as a function of $M_2$.}
\label{fig:rlof}
\end{figure}

\begin{figure*}
\centering
\includegraphics[width=0.9\textwidth]{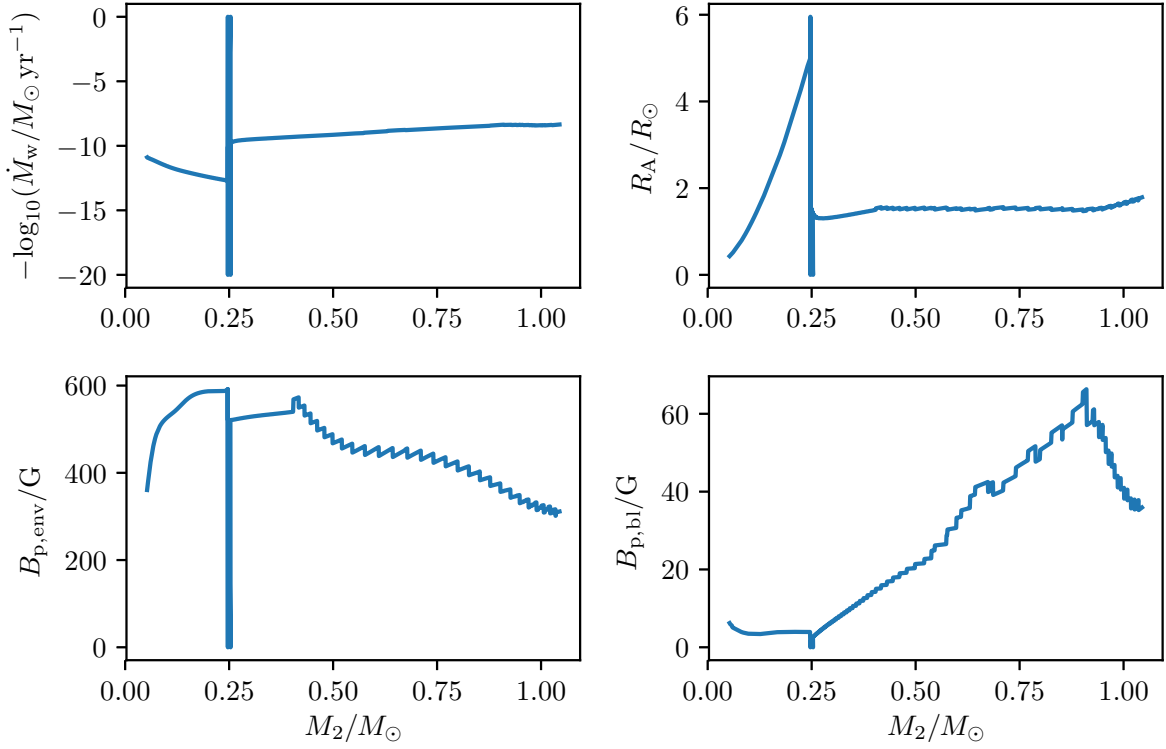}
\caption{Evolution of important terms in the DD model as a function of $M_2$ for the same system in Fig.~\ref{fig:pmfull}. The abrupt changes in the evolution of each parameter at $M_2\approx 0.25M_\odot=M_\mathrm{conv}$ are due to the interruption of the boundary layer dynamo.}
\label{fig:dd}
\end{figure*}
We use the tuple $(\alpha,\beta,\gamma) = (4.6,0.08,3.2)$ for the complete evolution of a CV with a donor of mass $M_2 = 1M_\odot$ and a WD of mass $M_1 = 0.83 M_\odot$. Fig.~\ref{fig:pmfull} shows the full evolutionary track of our CV in the $(M_2,P_\mathrm{orb})$ plane. The period gap is between $2.3$ and $3.4\,\mathrm{hr}$. However we emphasize that this is sensitive to the mass of the WD and so do not attempt to reproduce a precise gap yet. The period minimum is $P_\mathrm{orb, min}\:=\:82.89\,\mathrm{min}$ when $M_2=0.061M_\odot$. Fig.~\ref{fig:hr} is the evolution of the donor in an HR diagram {showing that, in order to evolve the system, we need a good implementation of a stellar EOS for temperatures below $\mathrm{log}_{10}(T/\mathrm{K})\lesssim3.5$} and Fig.~\ref{fig:rlof} shows how the mass ratio $q=M_2/M_1$, which in our figure is just a straight line with slope $1/M_1$ indicating complete non-conservative mass transfer, and the mass-transfer rate evolve with $M_2$. Fig.~\ref{fig:dd} shows the evolution of all the important terms in the DD model. There are important features to note.
\begin{enumerate}
    \item The Alfvén radius experiences a sudden increase near $M_\mathrm{conv}$ when the boundary layer dynamo stops operating. This can be explained by equation (\ref{ra}). Here $\Dot{M}_\mathrm{w}$ decreases abruptly at the upper end of the period gap leading to an increase in $R_\mathrm{A}$.
    \item The field $B_\mathrm{p,bl}$ attains a maximum quite far away from $M_\mathrm{conv}$, where the donor becomes fully convective. This contrasts with the result obtained by ZTB with their bipolytropic model in their fig.~4. 
\end{enumerate}

\subsection{Probability distributions}
\label{ss:pr}

\begin{figure}
\includegraphics[width=0.5\textwidth]{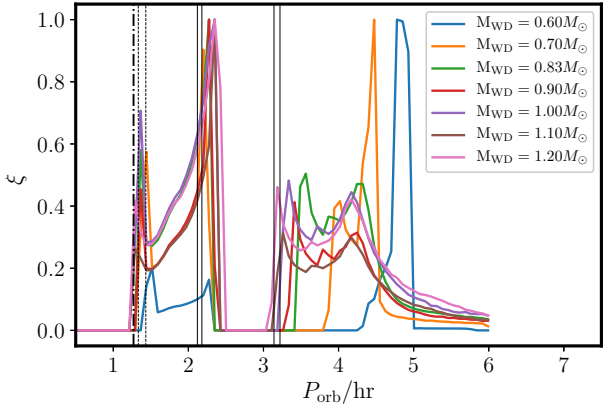}
\caption{The relative probability $\xi$ of a mass transferring system {existing} with a given $P_\mathrm{orb}$ for different WD masses $M_1$. The thin solid black lines are the lower and upper ends of the period gap given by $P_\mathrm{orb,pg,lower}\:=\:2.15\pm0.03\,\mathrm{hr}$ and $P_\mathrm{orb, pg,upper}\:=\:3.18\pm0.04\,\mathrm{hr}$ which we adopt from \protect\cite{Knigge2006}. The dotted black line is the period minimum spike $P_\mathrm{orb, min}\:\in\:[80,86]\,\mathrm{min}$ reported by \protect\cite{Gnsicke2009}. The thick black dash-dotted line is the minimum period $P_\mathrm{orb, min}\:=\:76.2\pm0.03\;\mathrm{min}$ reported by \protect\cite{Knigge2006}. }
\label{fig:P_distfull}
\end{figure}

\begin{figure}
\includegraphics[width=0.5\textwidth]{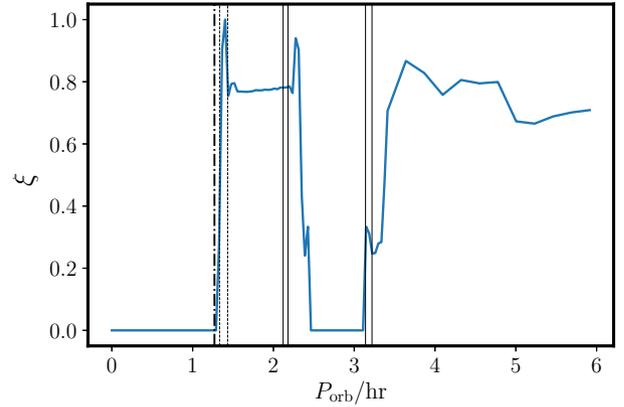}
\caption{The relative probability $\xi$ of a mass transferring system {existing} with a given $P_\mathrm{orb}$ scaled as per the WD distribution function from Fig.~6 of \protect\cite{Wijnen2015}. The vertical lines are the same as in Fig.~\ref{fig:P_distfull}.}
\label{fig:P_distscaled}
\end{figure}

We now estimate the probability of {having} a mass transferring system, where the Roche lobe radius $R_\mathrm{L}\geq R_\ast$, with a given $P_\mathrm{orb}$ in order to create a probability distribution histogram. We have shown in section \ref{pg} that $M_1$ affects the location of the period gap and the evolutionary track in general, so we evolve our $1M_\odot$ donor with a number of different WD accretors. Once we have its full evolution with time, we divide the orbital period space $P$ evenly in the range $P\in[0,6]\,\mathrm{hr}$ and define the probability $\xi$ of a system being found within a given bin $P\leq P_\mathrm{orb}< P+\mathrm{d}P$ as

\begin{center}
\begin{equation}
\label{prob}
\xi = 2\frac{t_\mathrm{max} - t_\mathrm{min}}{t_\mathrm{max} + t_\mathrm{min}},
\end{equation}
\end{center}
where $t_\mathrm{min}$ is the time when the system enters the $P$ bin and $t_\mathrm{max}$ is the time when the system leaves it. So that $\xi$ for a given $P$ bin is higher if the system stays in that bin for longer. Our results for a number of different WD masses are given in Fig.~\ref{fig:P_distfull}. We see that smaller accretor masses lead to wider period gaps and the systems spending more time away from the period gap.

To estimate a combined distribution, we generate a probability distribution histogram by scaling the individual trajectories in Fig.~\ref{fig:P_distfull} with the distribution of WDs in CVs \citep[see fig.~6 of][which uses the observed sample of \citealt{2011A&A...536A..42Z}]{Wijnen2015}. We use the WD number distribution from their black histogram. Our combined probability is shown in Fig.~\ref{fig:P_distscaled}. It can be seen that we reproduce the orbital period distribution of CVs of \cite{Knigge2006} and \cite[see their figs~4 and 2 respectively]{Gnsicke2009} taken from the catalogue of \cite{2003A&A...404..301R} and that the period gap and the period minimum spike discussed by \cite{Gnsicke2009} are reproduced quite well.

\subsection{Comparison with other work}
\label{ss:otherauth}
\begin{figure}
\includegraphics[width=0.5\textwidth]{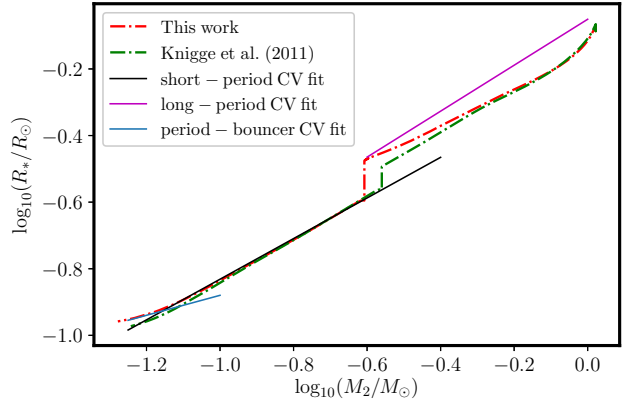}
\caption{$M$ to $R$ relationship of the donor star using our DD model and the model described by \protect\cite{Knigge2011} for a system with $M_2=1M_\odot$ and $M_1=0.83M_\odot$. The short-period CV fit (solid black line), the period-bouncer CV fit (blue line) and the long-period CV fit (magenta line) are the $M$ to $R$ best fits by \protect\cite{Knigge2011}.}
\label{fig:mr}
\end{figure}

We compare the $M$ to $R$ relationship of the donor from our model with that given by \cite{Knigge2011}, who used equation~(\ref{jdotbyjgr}) multiplied by 2.47 for angular momentum loss by gravitational radiation and an empirical formula for the angular momentum loss due to magnetic braking given by \cite{1983ApJ...275..713R},
\begin{center}
\begin{equation}
\label{jdotbyjrapp}
\Dot{J}_\mathrm{mb} \approx -3.8\times10^{{-30}} M_2 R_\odot^4 \left(\frac{R}{R_\odot}\right)^{\delta} \Omega^3 \; \mathrm{dyne\;cm},
\end{equation}
\end{center}
multiplied by 0.66, with $\delta=3$. Our results for a system with $M_2=1M_\odot$ and $M_1=0.83M_\odot$ are shown in Fig.~\ref{fig:mr}. We also plot the broken power law fit to the observed mass--radius distribution for CVs of \cite{Knigge2011}. We see that $M$ to $R$ fits of both the models are in excellent agreement with each other throughout most of their evolution. Our model excellently matches with the short-period CV fit (solid black line) and the period-bouncer CV fit (blue line). We see that the long-period CV fit (magenta line) does not match quite as well with either of the two model tracks. The reason for this is that long-period systems are populated by evolved donors as well as canonical CVs. These can follow completely different evolutionary tracks to ZACVs \citep[see the red and violet curves in fig.~1 of][]{Kalomeni2016}. Population synthesis studies have shown that at least half of the long-period CVs with $P_\mathrm{orb}\gtrsim6\,\mathrm{hr}$ have undergone some sort of nuclear evolution and some of these systems contribute to the observed present-day population of cataclysmic variables \citep[][]{Goliasch2015}. We believe that these systems have influenced long-period CV best fit made by \cite{Knigge2011}. 

\subsection{Comparison with observations}
\label{ss:obs}
\begin{figure}
\includegraphics[width=0.5\textwidth]{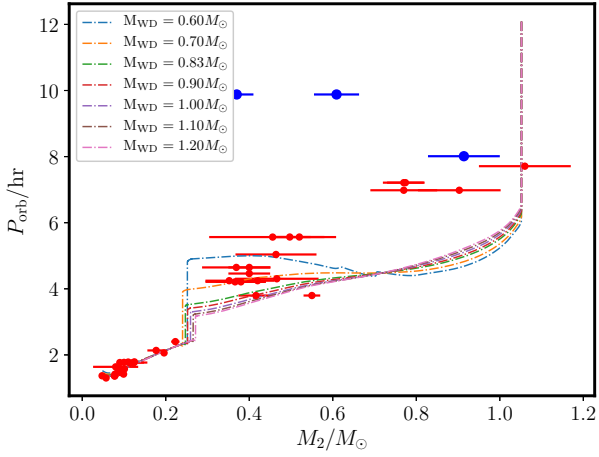}
\caption{Full evolutionary tracks for $M_2=1M_\odot$ for different WD masses $M_1$ plotted with observed CV data collected by \protect\cite{Ge2015}. The points in red are CVs with non-magnetic accretors whereas the points in blue are Intermediate polars.}
\label{fig:pm_obs1}
\end{figure}

\begin{figure}
\includegraphics[width=0.5\textwidth]{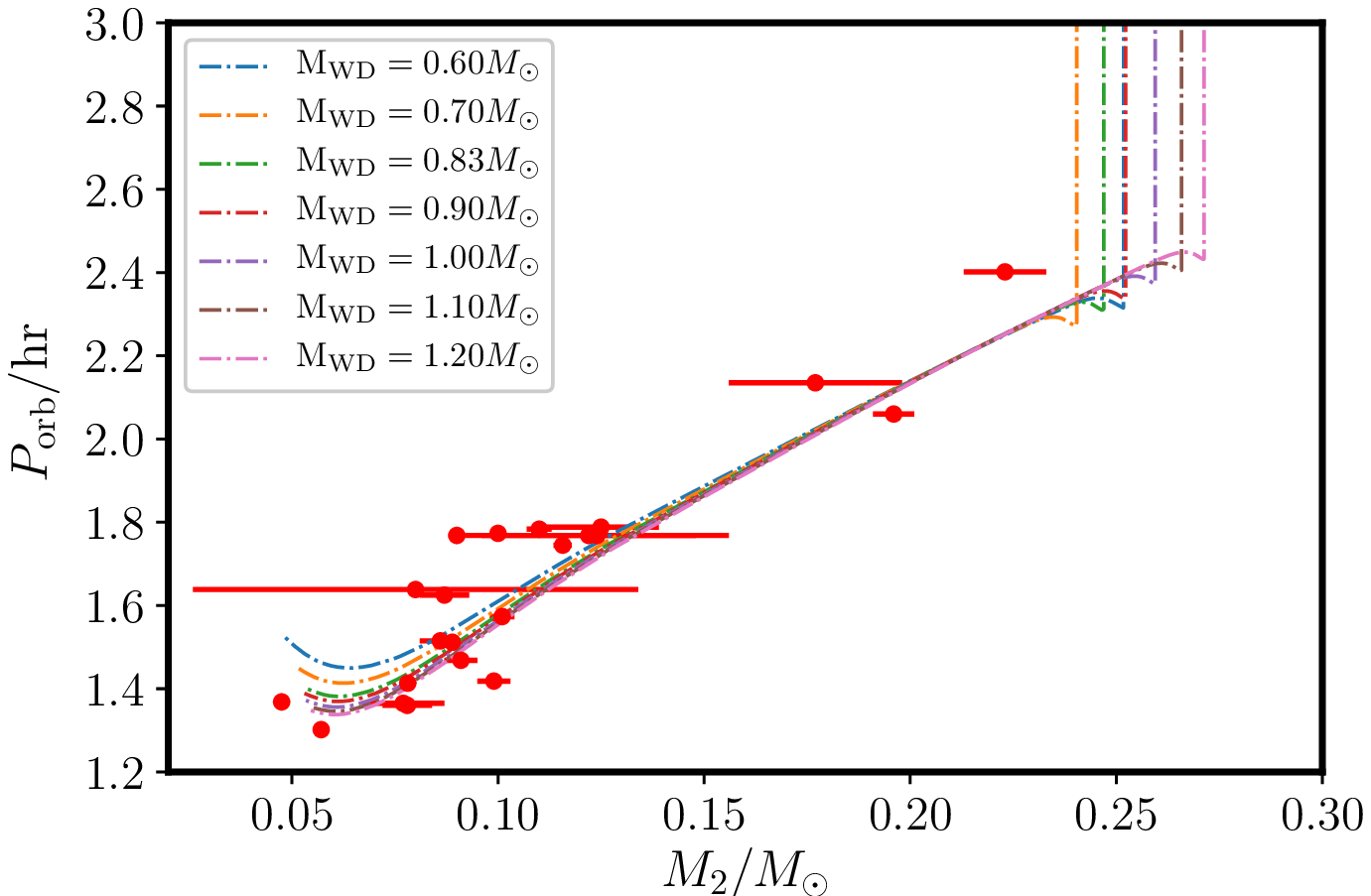}
\caption{Evolutionary tracks below the period gap for $M_2=1M_\odot$ for different WD masses $M_1$ plotted with observed CV data collected by \protect\cite{Ge2015}. }
\label{fig:pm_obs2}
\end{figure}

\begin{figure}
\includegraphics[width=0.5\textwidth]{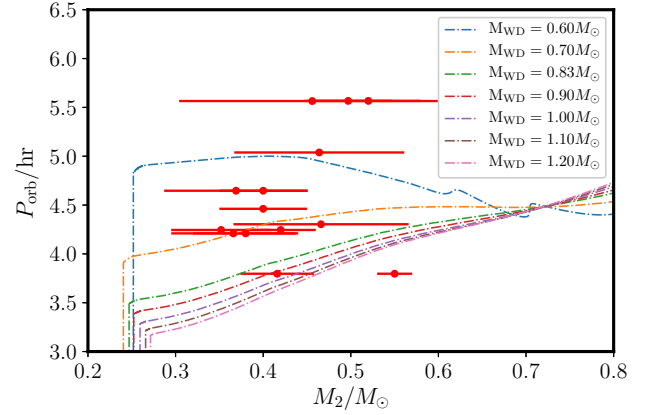}
\caption{Evolutionary tracks just above the period gap for $M_2=1M_\odot$ for different WD masses $M_1$ plotted with observed CV data collected by \protect\cite{Ge2015}. }
\label{fig:pm_obs3}
\end{figure}

We plot our evolutionary tracks in the $(M_2,P_\mathrm{orb})$ plane for different $M_1$ with observed CVs for which we have a robust estimate of $M_2$ and $P_\mathrm{orb}$. This is shown in Figs~\ref{fig:pm_obs1}, \ref{fig:pm_obs2} and \ref{fig:pm_obs3} where we use the data from table 5 of \cite{Ge2015}\footnote{Note that there are multiple observed parameters for the same systems in their table and we have plotted all of them. For instance, IP Peg has two measurements, one of which has donor mass $M_2=0.55M_\odot$ and does not lie on our set of trajectories.}. {The points in red are CVs with non-magnetic arrectors, whereas the points in blue (two measurements of AE Aqr and BT Mon) are Intermediate polars, in which mass-transfer to the accretor is affected by the WD magnetic field. We have not modelled these systems in this work.} We omitted plotting OV Boo\footnote{ OV Boo has $P_\mathrm{orb}=66.6\;\mathrm{min}$, less than the shortest CV period we consider here.}. We argue that this system is most likely an AM~CVn candidate. We also omit AM Her and DW UMa because reliable data are not available and U Sco with $P_\mathrm{orb}=1772\;\mathrm{min}>22\;\mathrm{hr}$, the bifurcation limit for our donor. We see that below the period gap (Fig.~\ref{fig:pm_obs2}) and right above the period gap (Fig.~\ref{fig:pm_obs3}) the observed data match extremely well with our our trajectories. However we see that systems with $P_\mathrm{orb}\gtrsim6\;\mathrm{hr}$ lie well off our trajectories. We again believe that these are systems where there has been some nuclear evolution of the donor star. That is these are systems with evolved secondaries and not the ZACVs which we have modelled here. 

\section{Conclusion}
\label{Conclusion}
Using an improved implementation of the equation of state module in the STARS code, we have come up with a unified model for the evolution of cataclysmic variables with a revised double dynamo model. With a set of three free parameters our model is able to not only explain the interrupted magnetic braking paradigm but also provide a mechanism for the extra angular momentum loss below the period gap. Our physically motivated expressions for the magnetic braking and its interruption when the donor becomes fully convective agrees well with the empirical formula by \cite{1983ApJ...275..713R} in the mass--radius relationship of the donor star. We show that the secular evolution of CVs is sensitive to the accretor mass and use the {mass} distribution of WDs in CVs to come up with a relative probability distribution of {having} a mass transferring system with a particular orbital period. We find that not only is the period gap reproduced well but the period minimum spike in the probability distribution is also seen. Comparing our model with observations we find good agreement between the two in short-period CVs and CVs right above the period gap. We argue that a substantial number of long-period CVs have evolved donors which we have not modelled in this work but shall consider in more detail in the future.

\section*{Acknowledgements}
AS thanks the Gates Cambridge Trust for his scholarship.  CAT thanks Churchill
College for his fellowship. AS also thanks Alex Hackett for insightful discussions in this field and Hongwei Ge for providing the observational data with which we compare our models.

\section*{Data availability}

No new data were generated or analysed in support of this research. The numerical code used to calculate the equation of state can be found at \url{https://github.com/ArnabSarkar3158/STARS-EOS-}. Any other numerical codes and related data generated during the work will
be available whenever required by the readers.

\bibliographystyle{mnras}
\bibliography{example} 




 \appendix

\section{Dependence of the secular evolution of the system on donor mass}
\label{app:1}
\begin{figure}
\includegraphics[width=0.5\textwidth]{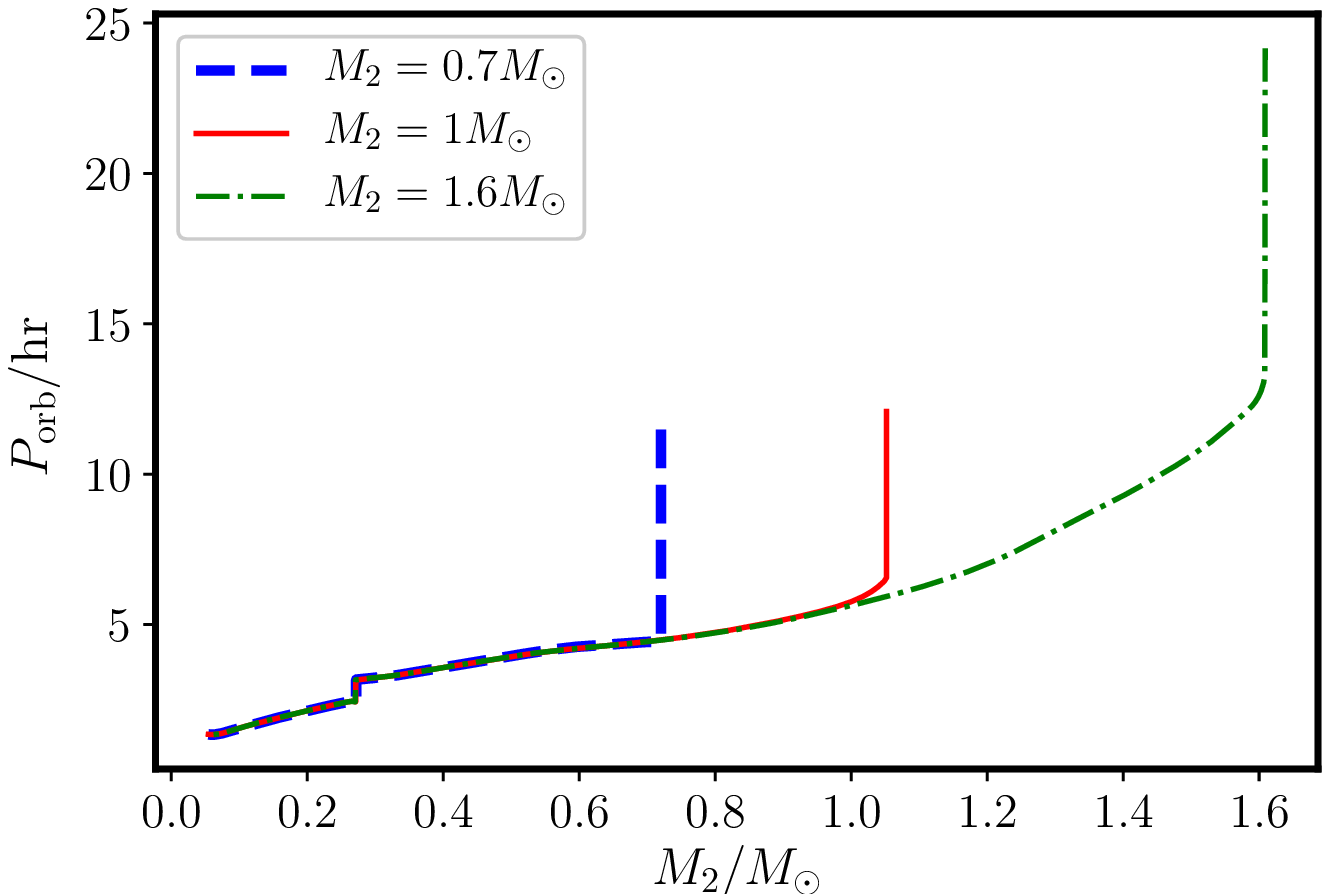}
\caption{Full evolutionary track of our CV model in the $(M_2,P_\mathrm{orb})$ plane for three different donor masses, where $(\alpha,\beta,\gamma) = (4.6,0.08,3.2)$ and $M_1=1.20M_\odot$.}
\label{fig:pm_m2}
\end{figure}

\begin{figure}
\includegraphics[width=0.5\textwidth]{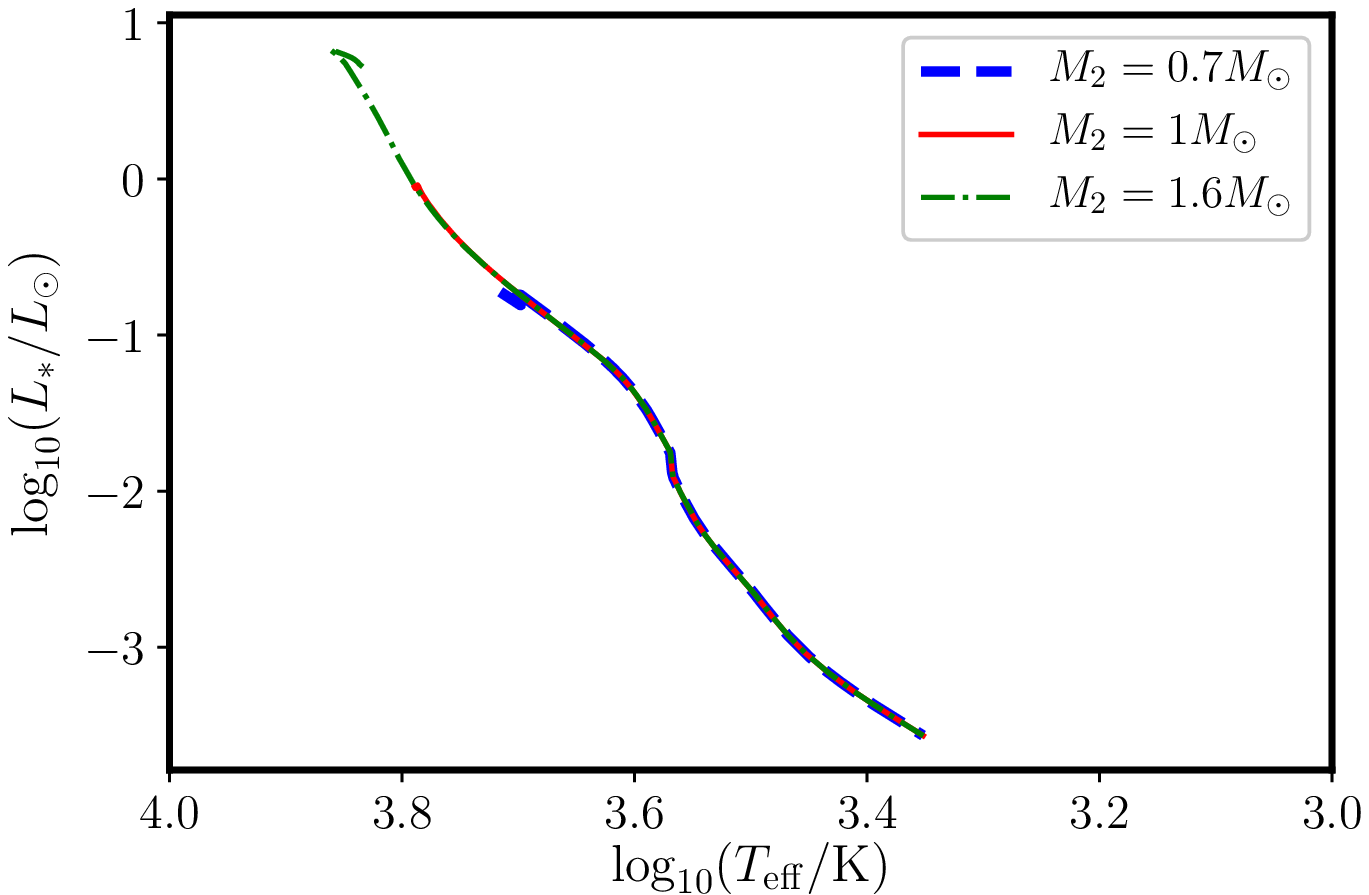}
\caption{The HR diagram evolution of the donor for three different donor masses, with the same system parameters as in Fig.~\ref{fig:pm_m2}.}
\label{fig:hr_m2}
\end{figure}

In order to confirm that the secular evolution of ZACVs is not dependent on the mass of the donor star $M_2$, we construct trajectories of systems with different $M_2$ and initial orbital period (such that none of the donors undergoes nuclear evolution) keeping all other parameters fixed. This is shown in Figs~\ref{fig:pm_m2} and \ref{fig:hr_m2} in the $(M_2,P_\mathrm{orb})$ plane and HR diagram. We see in Fig.~\ref{fig:pm_m2} that after Roche lobe overflow the trajectories catch up with each other and follow a common evolutionary track. Similarly in Fig.~\ref{fig:hr_m2} we see that the every donor eventually follows a common cooling curve in the HR diagram.


\bsp	
\label{lastpage}
\end{document}